\documentclass[12pt,preprint]{aastex}

\slugcomment{Not to appear in Nonlearned J., 45.}

\shorttitle{Effect of interstellar helium on the heliospheric
interface} \shortauthors{Izmodenov et al.}

\begin{document}


\title{Effects of interstellar and solar wind ionized helium on the
interaction of the solar wind with the local interstellar medium}


\author{Vlad Izmodenov\altaffilmark{1}, Yury G. Malama \altaffilmark{2},
George Gloeckler\altaffilmark{3} and Johannes Geiss
\altaffilmark{4}}

\altaffiltext{1}{Lomonosov Moscow State University, Department of
Aeromechanics and Gas Dynamics, Faculty of Mechanics and
Mathematics, Moscow, 119899, Russia; izmod@ipmnet.ru}
\altaffiltext{2}{Institute for Problems in Mechanics, Russian
Academy of Sciences, Prospekt Vernadskogo 101-1, Moscow, 119526,
Russia; malama@ipmnet.ru} \altaffiltext{3}{Department of Physics
and IPST, University of Maryland, College Park, MD 20742, USA;
gg10@umail.umd.edu} \altaffiltext{4}{International Space Science
Institute, Hallerstrasse 6, CH-3012 Bern, Switzerland;
geiss@issi.unibe.ch}


\begin{abstract}
The Sun is moving through a warm ($\sim$6500 K) and partly ionized
local interstellar cloud (LIC) with a velocity of $\sim$26 km/s.
Recent measurements of the ionization of the LIC (Wolff et al.,
1999) suggest that interstellar helium in the vicinity of the Sun
is 30-40 \% ionized, while interstellar hydrogen is less ionized.
Consequently, interstellar helium ions contribute up to 50\% of
the total dynamic pressure of the ionized interstellar component.
Up to now interstellar helium ions have been ignored in existing
models of the heliospheric interface. In this paper we present
results of a new model of the solar wind interaction with the
interstellar medium, which takes into account interstellar helium
ions. Using results of this model we find that the heliopause,
termination and bow shocks are closer to the Sun when compared to
the model results that ignore $He$ ions.  The influence of
interstellar helium ions is partially compensated by solar wind
alpha particles, which are taken into account in our new model as
well. Finally, we apply our new model to place constraints on the
plausible location of the termination shock.
\end{abstract}



\section{Introduction}

The solar wind interacts with the local interstellar cloud to form
the heliospheric interface, which separates the pristine
interstellar medium from the unperturbed solar wind.  The solar
wind meets the interstellar charged component at the heliopause
(HP), where the solar wind pressure balances the pressure of the
LIC. Since the solar wind is a supersonic flow, the heliospheric
termination shock (TS) should be formed to make the solar wind
subsonic before it reaches the heliopause.  Because the
interstellar flow is also supersonic (V$_{LIC}\sim$ 26 km/s,
T$_{LIC}$ $\sim$ 6500 K), a bow shock may be formed in the
interstellar medium. The idealized structure of the heliospheric
interface is shown in Figure 1.

Theoretical studies of the heliospheric interface began more than
four decades ago.
Recent models of the heliospheric interface take into account the
multi-component nature of both the LIC and the solar wind
The LIC consists of at least five components:
plasma (electrons and protons), hydrogen atoms (and other less
abundant atomic species), interstellar magnetic field, galactic
cosmic rays, and interstellar dust. The heliospheric plasma
includes solar wind particles (protons, electrons, alpha
particles, etc.), pickup ions and anomalous cosmic rays (ACRs),
which are pickup ions believed to be accelerated to high energies
at the termination shock.  To construct a realistic theoretical
description of the heliospheric interface, one needs to choose a
specific approach for each interstellar and heliospheric
components
(see, e.g.,  Baranov and Malama, 1993; Alexashov et al.,2000; Fahr
et al., 2000; Myasnikov et al., 2000; Izmodenov et al., 2003;  for
review, see also, Zank, 1999). Up to now, interstellar ionized
helium ions were ignored in the multi-component modeling of the
solar wind interaction with the LIC.  Recent measurements of
interstellar helium atoms (Witte et al., 2002) and interstellar He
pickup ions (Gloeckler and Geiss, 2002) inside the heliosphere as
well as of the interstellar helium ionization (Wolff et al., 1999)
allow us to estimate the number density of interstellar helium
ions to be 0.008-0.01 cm$^{-3}$. Current estimates of proton
number density in the LIC fall in the range of 0.04 - 0.07
cm$^{-3}$.  Since helium ions are four times heavier than protons
the dynamic pressure of the ionized helium component is comparable
to the dynamic pressure of the ionized hydrogen component.
Therefore, interstellar ionized helium cannot be ignored in the
modeling of the heliospheric interface. In this paper we present
results of our new model, which for the first time takes into
account interstellar ionized helium. Simultaneously with
interstellar ionized helium we took into account solar wind alpha
particles, which constitute 2.5 - 5 \% of the solar wind and,
therefore, produce 10 - 20 \% of the solar wind dynamic pressure.

\section{Model}

In this work we start with the global model of the heliospheric
interface developed  by the Moscow group (Baranov and Malama,
1993, Izmodenov et al., 1999; Alexashov et al., 2000; Myasnikov et
al., 2000; Izmodenov and Alexashov, 2003; Izmodenov et al., 2003;
see, also, for review Izmodenov, 2001) and introduce interstellar
ionized helium (He$^{+}$) and solar wind alpha particles
(He$^{++}$) into the model. We consider all plasma components
(electrons, protons, pickup ions, interstellar helium ions, and
solar wind alpha particles) as one-fluid with the total density
$\rho$ and bulk velocity v. This one-fluid approximation assumes
that all ionized components have the same temperature T. Although
this assumption cannot be made in the case of the solar wind, the
one-fluid model is based on mass, momentum and energy conservation
laws and predicts plasma bulk velocity and locations of the shocks
very well.

The plasma is quasineutral, i.e. $n_e = n_p+ n_{He^+}$ for the
interstellar plasma and $n_e = n_p+ 2 n_{He^{++}}$ for the solar
wind.  We ignore the magnetic field. While the interaction of
interstellar H atoms with protons by charge exchange is important,
for helium ions the process of charge exchange is negligible due
to small cross sections for charge exchange of helium atoms.
Hydrodynamic Euler equations for the charged component are solved
self-consistently with the kinetic equation for interstellar H
atom component.
Governing equations for the charged component are:
\begin{eqnarray}
\frac{\partial \rho }{\partial t}+ div (\rho \vec{v})=q_1,
\nonumber \\
\frac{\partial \rho \vec{V}}{\partial t}+ div (\rho
\vec{v}\vec{v}+ p \hat{I}) = \vec{q}_2 \\
 \frac{\partial
E}{\partial t} + div ([E+p]\vec{v}) = q_3 \nonumber
\end{eqnarray}
Here $\rho $ is the total density of the ionized component, $p$ is
the total pressure of the ionized component, $E=\rho (\varepsilon+
\vec{v}^2/2)$ is the total energy per unit volume, $\varepsilon =
\frac{p}{(\gamma-1)\rho}$ is the specific internal energy.
 The temperature of the plasma is determined from the equation of state
$p=2(n_p+ n_{He^+}) k T$ for the interstellar plasma and $p= (2
n_p+ 3 n_{He^{++}})k T$ for the solar wind.  To calculate sources
of mass, momentum and energy into the charged component due to
charge exchange of H atoms with protons and photo-ionization and
electron-impact-ionization processes we need to know $n_p$ in
addition to the total plasma density. (Expressions for the sources
 can be found, for example, in Izmodenov and Alexashov,
2003.)
 We solve the continuity
equations for He$^+$ in the interstellar medium and for alpha
particles in the solar wind. Then proton number density can be
calculated as $n_p = (\rho - m_{He} n_{He})/m_p$.  Here $n_{He}$
denotes the He$^+$ number density in the interstellar medium, and
He$^{++}$ the number density in the solar wind. The velocity
distribution of H atoms $f_{\rm H}(\vec{r}, \vec{w}_{\rm H}, t)$
is calculated from the linear kinetic equation introduced in
Baranov \& Malama (1993). The plasma and neutral components
interact mainly by charge exchange. However, photo-ionization,
solar gravitation, and radiation pressure, which are taken into
account in the governing equations, are important at small
heliocentric distances. Electron-impact ionization may be
important in the inner heliosheath, the region between the
termination shock and the heliopause.

\section{Boundary conditions}

The boundary conditions are the following.  At the Earth orbit we
assume that solar wind is spherically symmetric, which makes our
model axisymmetric, and we use IMP 8 data averaged over several
solar cycles for the solar wind parameters: $ n_{p,E}$ = 7.39
cm$^{-3}$, $V_{sw,E}$= 432 cm$^{-3}$.  The number density of solar
wind alpha particles is varied in our calculations from 0 \% to
4.5 \% of the solar wind proton number density.  The total density
and pressure of the solar wind at the inner boundary at 1 AU are
then:
\[
\rho_E = m_p n_{p,E}+ m_{He} n_{He^{++},E} \]
\[
 p_E= (2 n_{p,E}+ 3 n_{He^{++},E})k T_E.
\]

Among the interstellar parameters influencing the heliospheric
interface structure, the LIC velocity relative to the Sun and the
temperature of the local interstellar gas are now well established
by direct measurements of interstellar helium atoms with the GAS
instrument on Ulysses (Witte et al., 2002).  Unlike interstellar
hydrogen, atoms of interstellar helium penetrate the heliospheric
interface nearly undisturbed, because of negligible strength of
the coupling with protons due to the small cross sections of
elastic collisions and charge exchange. Based on these
measurements we take in this paper the temperature of the
interstellar gas to be 6500 K and the speed of the LIC relative to
the Sun as 26.4 km/s. The remaining three input parameters
required to calculate the heliospheric interface structure are the
number densities of interstellar protons, n$_{p,LIC}$, of
interstellar helium ions, $n_{He^{+},LIC}$, and of H atoms,
n$_{H,LIC}$.  To find good estimates of these important LIC
parameters we use (1) our measurements of the atomic H density at
the TS (=0.100 $\pm$ 0.008 cm$^{-3}$), (2) measurements of the LIC
atomic He density (= 0.015 $\pm$ 0.002 cm$^{-3}$) (Gloeckler and
Geiss, 2001; Witte, private communication), (3) the standard
universal ratio of the total H to He,
(n$_{p,LIC}$+n$_{H,LIC}$)/(n$_{He^+,LIC}$+n$_{He,LIC}$) = 10, and
(4) measurements of the local interstellar helium ionization
fraction, n$_{He^+,LIC}$/(n$_{He^+,LIC}$+n$_{He,LIC}$) = 0.35
$\pm$ 0.05 (Wolff et al., 1999). Previously, a similar method was
used to determine interstellar H atom and proton number densities
by Lallement (1996), Gloeckler et al. (1997), Izmodenov et al.
(2003). With these constraints we find that the heliospheric
interface model with n$_{H,LIC}$ = 0.18 $\pm$ 0.02 cm$^{-3}$,
n$_{p,LIC}$ = 0.06 $\pm$ 0.015 cm$^{-3}$ and n$_{He^+,LIC}$ =
0.009  cm$^{-3}$ provides the best fit to SWICS Ulysses pickup
hydrogen data.  The interstellar hydrogen ionization fraction
derived from our results is in agreement with recent calculations
of the photo-ionization of interstellar matter within 5 pc of the
Sun (Slavin and Frisch, 2002).

 The total density and pressure of the interstellar gas are calculated to be:
\[
\rho_{LIC} = m_p n_{p,LIC} +m_{He} n_{He^{+},LIC},
\]
\[
p_{LIC} = 2( n_{p,LIC}+ n_{He^{++},LIC})k T_{LIC}.
\]

\section{Results}

To evaluate possible effects of both interstellar ions of helium
and solar wind alpha particles we performed parametric model
calculations with eight different sets of boundary conditions
given in Table 1.  Calculated locations in the upwind direction of
the termination shock, the heliopause and the bow shock are given
for each model in the last three columns of Table 1 respectively.
In the first six models we assume that n$_{p,LIC}$ = 0.06
cm$^{-3}$ and n$_{H,LIC}$ = 0.18 cm$^{-3}$. Model 1 does not
include either interstellar ionized helium or solar wind alpha
particles.  Effects of interstellar ionized helium can be seen by
comparing results of model 1, with no ionized interstellar He, and
model 2, where 37.5 \% ionization of interstellar helium is
assumed. Interstellar helium ions increase the interstellar
dynamic pressure by 60 \% and the interstellar thermal pressure by
15 \%  in model 2 as compared with model 1:
\[
\frac{(\rho v^2)_{LIC, model 2}}{({\rho v^2})_{LIC, model 1}} =
\frac{n_{LIC,p}+4 n_{LIC,HE^+}}{n_{LIC,p}}= 1.6,
\]
\[
\frac{p_{LIC, model 2}}{p_{LIC, model 1}} =
\frac{n_{LIC,p}+n_{LIC,HE^+}}{n_{LIC,p}}= 1.15.
\]
This additional interstellar pressure pushes the bow shock, the
heliopause and the termination shock towards the Sun, from the
dashed to the solid curves of Figure 1.  In model 2 the heliopause
is $\sim$20 AU, and the termination shock  $\sim$7 AU closer to
the Sun as compared with model 1. The influence on the bow shock
location is even stronger.  The BS is $\sim$50 AU closer to the
Sun as compared with model 1.  This strong displacement of the bow
shock toward the Sun is also connected with the fact that the Mach
number is larger when ionized helium is taken into account.
Indeed,
\[
M=\frac{V}{\sqrt{\gamma P/\rho}}= V
\sqrt{\frac{n_{p,LIC}+4n_{He^{+},LIC}}{n_{p,LIC}+
n_{He^{+},LIC}}\frac{m_p}{2 k_B T_{LIC}}}= 2.3,
\]
as compared to M=1.97 for model 1. Here $m_p$, $k_B$ are the
proton mass and the Boltzmann's constant, respectively.  The
plasma compression at the bow shock is 1.45 for model 2 and 1.22
for model 1.  Higher compression of the interstellar plasma at the
bow shock and the corresponding reduction of the size of the outer
heliosheath - the distance between the bow shock and the
heliopause - make the optical depth for interstellar H atoms in
the interface about the same for all models 1-6.  The resulting
filtration of interstellar hydrogen atoms in the interface is
therefore about the same in all of the first six models. The
effects of the solar wind alpha particles are seen from comparison
of results of model 1 with model 3. Influence of solar wind alpha
particles on the locations of the heliopause and shocks is
opposite to the influence of interstellar helium ions discussed
above.  Since solar wind alpha particles constitute only 10 - 18
\% of the solar wind dynamic pressure, their influence is less
pronounced.  The heliopause and the termination shock move out by
$\sim$6 AU and $\sim$5 AU from the Sun, respectively.

The net effect of both the interstellar ionized helium and solar
wind alpha particles is seen by comparing models 4-6.  Model 5
corresponds to 37.5 \% of the interstellar helium ionization, and
2.5 \% the solar wind alpha particles abundance. The influence of
interstellar helium ions on the locations of HP, BS, and TS is
stronger than the influence of the solar wind alpha particle
component.  The heliopause is located $\sim$ 13 AU closer to the
Sun in model 5 than model 1. We note with interest that Gurnett et
al. (1993, 1995) analyzing heliospheric radio emission events of
1983-84 and 1992-94 at plasma cutoff frequency, fp = 2.2 to 2.8
kHz detected by Voyagers estimated the average distance to the
heliopause to be ~158 AU, close the HP distance we find for model
5.  The bow shock is closer to the Sun by $\sim$40 AU.  At the
same time the termination shock location is only $\sim$2 AU closer
to the Sun. For smaller interstellar He ionization (model 4) or a
higher abundance of solar wind alpha particles (model 6) the
termination shock is  4-5 AU further from the Sun as compared to
model 5. To estimate the influence of the interstellar ionized
helium component in the case of smaller hydrogen ionization, we
took n$_{p,LIC}$ =0.04 cm$^{-3}$ and n$_{H,LIC}$ = 0.20 cm$^{-3}$
in models 7 - 8. Expectedly, the effect on the locations of the
heliopause and termination shock is about the same as previously
(see table 1).  The bow shock is $\sim$50 AU closer in model 8 as
compared to model 7.

\section{Implications on the TS Location}

Using our model and boundary conditions described above, we
performed parametric studies by varying the interstellar proton
and hydrogen atom number densities in the ranges of 0.03-0.1
cm$^{-3}$ and 0.16-0.2 cm$^{-3}$, respectively.  The interstellar
helium ion number density was calculated by using an interstellar
helium atom number density of 0.015 cm$^{-3}$ and 10 for the
interstellar H/He ratio.  Figure 2 shows results of our
calculations.  Displayed are contour isolines of (1) the neutral
hydrogen density at the TS, (2) the LIC helium ionization
fraction, and (3) the termination shock location in the upwind
direction on a (n$_{H,LIC}$,n$_{p,LIC}$)-coordinate plane.  Dashed
areas show (a) the n$_{H,TS}$ range of 0.095 - 0.105 cm$^{-3}$,
which corresponds to recent Ulysses determination (Gloeckler and
Geiss, 2002); (b) the interstellar helium ionization fraction
range of 0.3 -0.4 derived from  line-of-sight EUVE measurements
toward white draft stars in the LSIM (Wolff et al., 1999). The
intersection of the two dashed areas gives a likely range of
interstellar proton and atomic hydrogen number densities
compatible with observations. Using long-term averages of IMP 8
solar wind parameters places the average termination shock
location at more than 90 AU in the upwind direction and more than
95 AU in the direction of Voyager 1 for all pairs of
(n$_{H,LIC}$,n$_{p,LIC}$) in this doubly-dashed area. Solar-cycle
variations of the solar wind ram pressure lead on average to a 7
to 8 AU deviation of the termination shock distance around its
mean value (Izmodenov, et al., 2003). In 2002 the termination
shock had its minimal location (Izmodenov et al., 2003), which
according to our model calculations should not have been less than
87-88 AU under average solar wind conditions at that time of the
solar cycle.  Based on measurements of low energy particle fluxes,
spectra and composition by the Voyager-1/LECP instrument, and of
indirect determination of the solar wind speed using particle
anisotropy measurements Krimigis et al. (2003) reported the
probable crossing of the termination shock by Voyager-1 at ~85 AU
in the summer of 2002 and return to the TS upstream region about
six months later. Temporary and probably localized excursions of
the termination shock inward by a few AU beyond our minimum value
cannot be ruled out by our calculations, since they could result
from an anomalously low solar wind ram pressure and possibly other
causes. However, should future measurements show that the TS
location is consistently less than what we calculate here, then a
revision of the LIC He ionization to higher values and or a
stronger local interstellar magnetic field may be required.

\section{Summary and conclusions}

We studied the influence of the interstellar ionized helium
component on the heliospheric interface for the first time.  This
component may create up to 50 \% of total dynamic pressure of the
interstellar ionized component.  It is shown the heliopause,
termination and interstellar bow shocks are closer to the Sun when
influence of interstellar helium ions is taken into account.  This
effect is partially compensated by additional solar wind alpha
particle pressure that we also took into account in our model. The
net result is as follows:  the heliopause, termination and bow
shocks are closer to the Sun by $\sim$12 AU, $\sim$ 2 AU, $\sim$
30 AU, respectively in the model taking into account both
interstellar helium ions and solar wind alpha particles (model 5)
as compared to the model ignoring these ionized helium components
(model 1).  We also found that both interstellar ionized helium
and solar wind alpha particles do not influence the filtration of
the interstellar H atoms through the heliospheric interface.

We use our model to determine a plausible range of
(n$_{H,LIC}$,n$_{p,LIC}$) compatible with (1) $n_{H,TS} = 0.1 \pm
0.05$ cm$^{-3}$ determined by Ulysses/SWICS, (2) ionization of
interstellar helium 0.35 $\pm$0.05.  Using our model we found that
the lower limit (1-$\sigma$ ) of the termination shock location in
the direction of Voyager-1 is ~88 AU.  While temporary and
localized motions of the termination shock position as close as 85
AU cannot be ruled out, definitive experimental determination of
the average termination shock location in the near future would
place a firm additional constraint on the possible ranges of
interstellar parameters.

{\it Acknowledgements.} We thank the International Space Science
Institute (ISSI) staff for their hospitality during our visit to
ISSI where discussions leading to results of this publication were
initiated. This work was supported in part by the International
Space Science Institute in Bern, INTAS grant 2001-0270, RFBR
grants 01-02-17551, 01-01-00759, and NASA/Caltech grant NAG5-6912
and NASA/JPL contract 955460.

\clearpage
\begin{table}[t]
 \caption{Sets of model parameters and locations of the TS, HP and BS in the upwind direction}
\begin{tabular}{llllllll}
\hline
\\
\# &  $n_{H,LIC}$ & $n_{p,LIC}$ & $\frac{n_{\alpha,sw}}{n_{e,sw}}$ & $\chi_{He}$\tablenotemark{a} & R(TS) & R(HP) & R(BS)\\
\hline
      &    cm$^{-3}$ &   cm$^{-3}$ & \%                       &                        &  AU &AU &AU \\
\hline
1     & 0.18         & 0.06        & 0   & 0     & 95.6 & 170  & 320  \\
2     & 0.18         & 0.06        & 0   & 0.375 & 88.7 & 152  & 270  \\
3     & 0.18         & 0.06        & 2.5 & 0     &100.7 & 176  & 330  \\
4     & 0.18         & 0.06        & 2.5 & 0.150 & 97.5 & 168  & 310  \\
5     & 0.18         & 0.06        & 2.5 & 0.375 & 93.3 & 157  & 283  \\
6     & 0.18         & 0.06        & 4.5 & 0.375 & 97.0 & 166  & 291  \\
\hline
7     & 0.20         & 0.04        & 0   & 0     & 95.0   & 183  & 340  \\
8     & 0.20         & 0.04        & 2.5 & 0.375 & 93.0   & 171 & 290  \\
\hline
\end{tabular}
\tablenotetext{a}{$\chi_{He} =HeII/(HeI+HeII)$}
\end{table}


\clearpage
\begin{figure}
\plotone{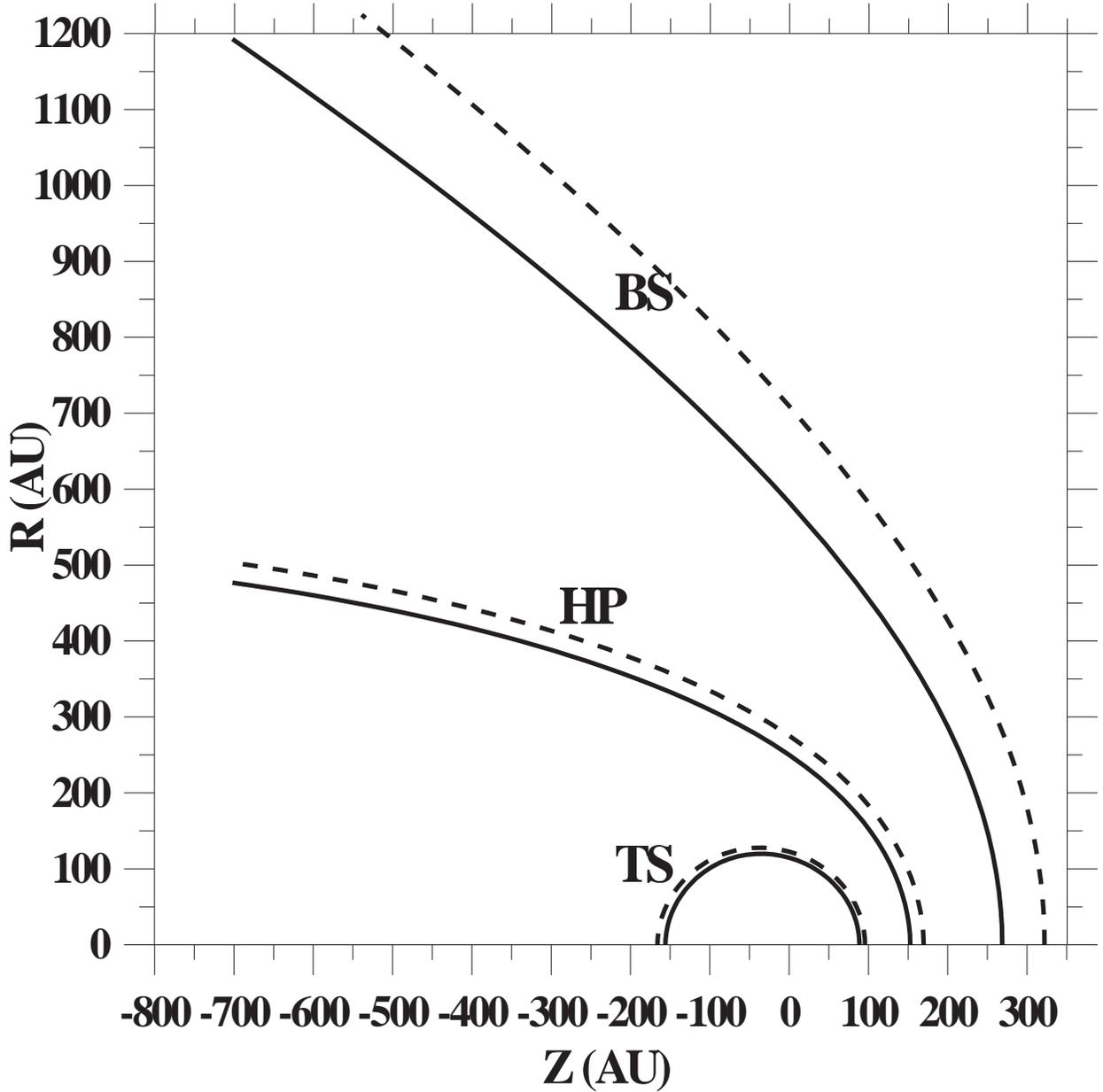} \caption{Sketched is the idealized structure of
the heliospheric interface (the region of interaction of the solar
wind with the LIC) based on results of numerical modeling. We used
the following interstellar parameters: (a) atomic hydrogen number
density (= 0.18 cm$^{-3}$), (b) proton number density (= 0.06
cm$^{-3}$), (c) gas temperature (= 6500 K), and (d) gas speed
(relative to the Sun) (= 26.4 km/s), and average solar wind
parameters: (e) solar wind density at 1 AU (= 7.39 cm$^{-3}$), and
(f) speed (= 432 km/s).  Discussion of, and references for the
chosen parameters are given in the text. Dashed curves correspond
to model 1, solid curves correspond to model 2 (see, Table 1).
\label{fig1}}
\end{figure}

\clearpage

\begin{figure}
\epsscale{.60}
\plotone{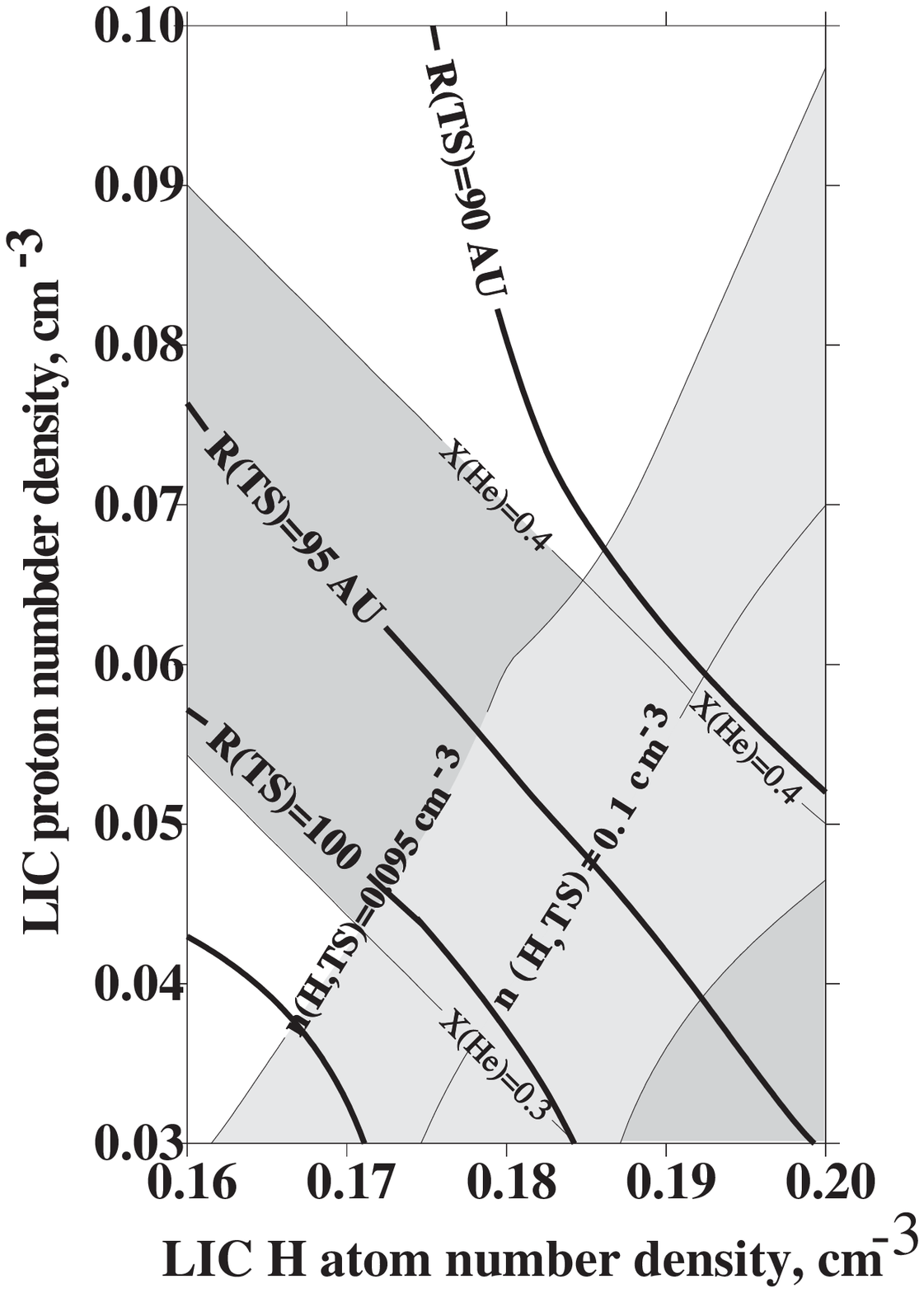} \caption{Contour plots of the interstellar H atom
number density at the termination shock, the LIC helium ionization
fraction, and the termination shock location in the upwind
direction.\label{fig2}}
\end{figure}


\begin{thebibliography}{}
\bibitem[Anders and Grevesse (1988)]{anders_grevesse1988} Anders, E., and Grevesse, N.,
{\it Geochim. Cosmochim. Acta} 53, 1197, 1988.
\bibitem[Alexashov(2000)]{alexash2000} Alexashov, D.~B.,
Baranov, V.~B., Barsky, E. V., and Myasnikov, A. V, An
Axisymmetric Magnetohydrodynamic Model for the Interaction of the
Solar Wind with the Local Interstellar Medium,
{\it Astronomy Letters} 26, 743-749, 2000.
\bibitem[Baranov and Malama (1993)]{bm93} Baranov, V.~B.,
Malama, Y.~G., Model of the solar wind interaction with the local
interstellar medium - Numerical solution of self-consistent problem,
{\it J. Geophys. Res.} 98, pp. 15,157-15,163, 1993.
\bibitem[Fahr et al. (2000)]{fahr2000} Fahr, H.-J., Kausch, T., Scherer, H.,
A 5-fluid hydrodynamic approach to model the solar
system-interstellar medium interaction,
Astron. Astrophys. 357, 268-282, 2000.
\bibitem[Gloeckler et al. (1997)]{gl97}Gloeckler, G.,
Fisk, L. A. Geiss J.,
Anomalously small magnetic field in the local interstellar cloud,
{\it Nature}, 386, 374-377, 1997.
\bibitem[Gloeckler and Geiss (2001)]{gl01}Gloeckler, G., and
Geiss, J.,
Joint SOHO/ACE workshop "Solar and Galactic
Composition". Ed. Wimmer-Schweingruber.
AIP Conference proceedings vol. 598 , 281, 2002.
\bibitem[Gurnett et al. (1993)]{gurnett93} Gurnett, D. A., Kurth, W. S., Allendorf, S. C., Poynter, R.
L.,
 Radio emission from the heliopause triggered by an
interplanetary shock
Science, 262, 199-203 (1993).
\bibitem[Gurnett and Kurth (1995)]{gurnett95} Gurnett, D. A., and Kurth, W. S.
Heliospheric 2-3 kHz radio
emissions and their relationship to large Forbush decreases.
Adv. Space Res. 9, 279-290 (1995)
\bibitem[Izmodenov et al. (1999)]{iglgbm99}Izmodenov, V.~V., Geiss, J., Lallement, R., Gloeckler, G.,
Baranov, V.~B., and Malama, Yu.~G.,
Filtration of interstellar
hydrogen in the two-shock heliospheric interface: inferences on
the LIC electron density,
J. Geophys. Res. 104, 4731-4741, 1999.
\bibitem[Izmodenov (2001)]{izmod01}Izmodenov, V.~V.,
Interstellar atoms in the heliospheric interface,
in Proceedings of COSPAR Colloquium on The Outer Heliosphere, Eds.
K. Scherer, et al., 23-32, 2001.
\bibitem[Izmodenov and Alexashov (2003)]{ia2003}Izmodenov, V., and, D. Alexashov,
A Model for the tail region of the heliospheric interface,
Astronomy Letters, Vol. 29, No. 1, pp. 58-63, 2003.
\bibitem[Izmodenov and Malama (2003)]{im2003}Izmodenov, V.~V., and Malama, Yu.~G.,
Variation of interstellar H atom filtration at the entrance to the
heliosphere: the solar cycle effect,
Adv. Space Res.,accepted, 2003.
\bibitem[Izmodenov et al. (2003)]{igm2003}Izmodenov, V.~V., Gloeckler, G., and Malama, Yu.~G.,
When Voyager 1 and 2 will encounter the termination shock?,
Geophys. Res. Letters, 2003.
\bibitem[Krimigis et al.(2003)]{krim2003} Krimigis, et al.,
AGU-EGS Assembly, Nice, 2003.
\bibitem[Lallement (1996)]{lall96} Lallement, R.,
Relations
Between ISM Inside and Outside the Heliosphere,
Space Science Reviews, 78, 361-374 (1996).
\bibitem[Myasnikov et al.(2000)]{myas2000} Myasnikov, A.V.,
Izmodenov, V.V., Alexashov, D. B., Chalov, S. V.,
Self-consistent
model of the solar wind interaction with two-component circumsolar
interstellar cloud: Mutual influence of thermal plasma and
galactic cosmic rays,
\jgr 105, 5179-5188, 2000.
\bibitem[Slavin and Frisch (2002)]{slavin2002} Slavin, J. D., and
Frisch, P.C.,
The Ionization of Nearby Interstellar Gas,
Astrophys. J. 565, 364-379, 2002.
\bibitem[Stone (2001)]{stone2001} Stone, E.C.,
News from the Edge of interstellar space,
{\it Science} 293, 55-56, 2001.
\bibitem[Witte et al. (1996)]{witte96} Witte, M., Banaszkiewicz, M.,
and Rosenbauer, H.,
Recent Results on the Parameters of the
Interstellar Helium from the Ulysses/Gas Experiment,
{\it Space
Science Reviews}, 78, 289-296,1996.
\bibitem[Wolff et al. (1999)]{wolff99} Wolff, B., Koester, D.,
and Lallement, R.,
Evidence for an ionization gradient in the
local interstellar medium: EUVE observations of white dwarfs,
{\it Astron. Astrophys.}, 346, 969-978,1999.
\bibitem[Zank (1999)]{zank99}Zank, G.,
Interaction of the solar wind
with the local interstellar medium: a theoretical perspective,
Space Science Reviews, 89, 413-688 (1999).
\end{thebibliography}
\end{document}